# GPU-based Fast Low-dose Cone Beam CT Reconstruction via Total Variation

**Xun Jia**
*Department of Radiation Oncology, University of California San Diego, La Jolla, CA 92037-0843*
**Yifei Lou**
*Department of Mathematics, University of California Los Angeles, Los Angeles, CA 90095*
**John Lewis, Ruijiang Li, Xuejun Gu, Chunhua Men, William Y. Song, and Steve B. Jiang[a]**
*Department of Radiation Oncology, University of California San Diego, La Jolla, CA 92037-0843*

**Purpose**: Cone-beam CT (CBCT) has been widely used in image guided radiation therapy (IGRT) to acquire updated volumetric anatomical information before treatment fractions for accurate patient alignment purpose. However, the excessive x-ray imaging dose from serial CBCT scans raises a clinical concern in most IGRT procedures. The excessive imaging dose can be effectively reduced by reducing the number of x-ray projections and/or lowering mAs levels in a CBCT scan. It is the goal of this work to develop a fast GPU-based algorithm to reconstruct high quality CBCT images from undersampled and noisy projection data so as to lower the imaging dose.

**Methods:** The CBCT is reconstructed by minimizing an energy functional consisting of a data fidelity term and a total variation regularization term. We developed a GPU-friendly version of the forward-backward splitting algorithm to solve this model. A multi-grid technique is also employed. We test our CBCT reconstruction algorithm on a digital NCAT phantom and a head-and-neck patient case. The performance under low mAs is also validated using a physical Catphan phantom and a head-and-neck Rando phantom.

**Results:** It is found that 40 x-ray projections are sufficient to reconstruct CBCT images with satisfactory quality for IGRT patient alignment purpose. Phantom experiments indicated that CBCT images can be successfully reconstructed with our algorithm under as low as 0.1 mAs/projection level. Comparing with currently widely used full-fan head-and-neck scanning protocol of about 360 projections with 0.4 mAs/projection, it is estimated that an overall 36 times dose reduction has been achieved with our algorithm. Moreover, the reconstruction time is about 130 sec on an NVIDIA Tesla C1060 GPU card, which is estimated ~100 times faster than similar iterative reconstruction approaches.

---

[a] Electronic mail: sbjiang@ucsd.edu

**Conclusions:** The developed GPU-based CBCT reconstruction algorithm is capable of lowering imaging dose considerably. The high computational efficiency makes this iterative CBCT reconstruction approach feasible in real clinical environments.

45  **I. INTRODUCTION**

Cone Beam Computed Tomography (CBCT) plays an important role in cancer radiotherapy. In CBCT, the patient's volumetric information is retrieved from its x-ray projections in cone beam geometry along a number of directions. In the past, the CBCT
50  reconstruction algorithms have undergone significant advances. Successful algorithms include the well known FDK algorithm[1] and a large number of iterative reconstruction algorithms, such as EM[2] and ART/SART[3-4], to name a few.

Among its various applications, CBCT is particularly convenient for accurate patient setup in image guided radiation therapy (IGRT). Yet, the high imaging dose to healthy
55  organs (a few cGy per scan)[5-7] in CBCT scans is a clinical concern, especially when CBCT is performed on a daily basis before each treatment fraction. The imaging dose in CBCT can be reduced by reducing the number of x-ray projections and/or lowering mAs levels (tube current and pulse duration). In this approach, however, the consequent CBCT images reconstructed with conventional FDK algorithms are highly degraded due to
60  insufficient and noisy projections. It is therefore desirable to develop new techniques to perform CBCT reconstructions with a greatly reduced number of noisy projections, while maintaining high image quality.

Reconstructing CBCT images from only few number of x-ray projections for IGRT is feasible, though quite difficult. First, for the purpose of patient setup before a
65  radiotherapy treatment, it is adequate to only retrieve major anatomic features from a CBCT image, such as tumor location and shape. This fact considerably reduces the complexity of the reconstruction problem. Second, recent development in mathematics provides us new perspectives of solving problems of this kind. Recently, a burst of research in compressed sensing[8-13] have demonstrated the feasibility of recovering signals
70  from incomplete measurements through optimization methods in many different mathematical situations. In particular, Total Variation (TV) method has presented its tremendous power in CT reconstruction problems, where the CT images can be successfully reconstructed by minimizing the TV semi-norm under the constraints posed by the x-ray projections along a few directions[14-15]. This approach has also been
75  extensively applied into many other imaging modalities[16-19] and its efficacy has been improved by combining with other techniques, such as incorporating prior information[20-21].

Despite the great power of TV method and its promising applications in the problem of CBCT reconstruction, the computation is very time-consuming due to the high
80  complexity of currently available algorithms and the extremely large data size involved in real clinical problems. It usually takes several hours or even longer for current TV-based reconstruction approaches to produce a decent CBCT image. This fact prevents them from practical applications in real clinical environments. Generally speaking, the reconstruction process can be sped up by utilizing a better optimization algorithm and a
85  more powerful computational hardware. Accompanied with the recent development of compressed sensing, a number of efficient minimization algorithms targeting at problems of this type have been invented such as graph cut[22-23], TV-operator splitting[24-25], and Bregman iterations[26-27]. Meanwhile, general purpose graphic processing units (GPUs) have offered us a promising prospect of increasing efficiencies of heavy duty tasks in radiotherapy, such as CBCT FDK reconstruction[28-32], deformable image registration[30, 33-34], dose calculation[35-38], and treatment plan optimization[39]. Taking advantages of these developments, the computation efficiency of TV-based CBCT reconstruction is expected to be enhanced considerably.

We have recently reported some preliminary results on the development of a GPU-based CBCT reconstruction algorithm in a short letter[40], where we have reconstructed CBCT images by minimizing an energy functional consisting of a TV regularization term and a data fidelity term posed by the x-ray projections. By developing new minimization algorithms with mathematical structures suitable for GPU parallelization, we can take advantage of the massive computing power of GPU to dramatically improve the efficiency of the TV-based CBCT reconstruction. In this paper, we will present a comprehensive description and validation of our algorithm.

## II. METHODS

### A. Model

Let us consider a patient volumetric image represented by a function $f(x, y, z)$, where $(x, y, z) \in \mathbf{R}^3$ is a vector in 3-dimensional Euclidean space. A projection operator $P^\theta$ maps $f(x, y, z)$ into another function on an x-ray imager plane along a projection angle $\theta$:

$$P^\theta [f(x, y, z)](u, v) = \int_0^{L(u,v)} dl \, f(x_S + n_1 l, y_S + n_2 l, z_S + n_3 l), \quad (1)$$

where $(x_S, y_S, z_S)$ is the coordinate of the x-ray source S and $(u, v)$ is the coordinate for point P on the imager, $\mathbf{n} = (n_1, n_2, n_3)$ being a unit vector along the direction SP. Fig. 1 illustrates the geometry. The upper integration limit $L(u, v)$ is the distance from source S to point P. Denote the observed projection image at the angle $\theta$ by $Y^\theta(u, v)$. A CBCT reconstruction problem is formulated as to retrieve the volumetric image function $f(x, y, z)$ based on the observation of $Y^\theta(u, v)$ at various angles given the projection mapping in Eq. (1).

In this paper, we attempt to reconstruct the CBCT image by minimizing an energy functional consisting of a data fidelity term and a regularization term:

$$f = \text{argmin } E_1[f] + \mu E_2[f], \text{ s.t. } f(x, y, z) \geq 0 \text{ for } \forall \, (x, y, z) \in \mathbf{R}^3, \quad (2)$$

where the energy functionals are $E_1[f] = \frac{1}{V} \|\nabla f\|_1$ and $E_2[f] = \frac{1}{N_\theta A} \sum_\theta \|P^\theta[f] - Y^\theta\|_2^2$. Here $V$ is the volume in which the CBCT image is reconstructed. $N_\theta$ is the number of projection used and $A$ is the projection area on each x-ray imager. $\|...\|_p$ denotes $l_p$-norm of functions. In Eq. (2), the data fidelity term $E_2[f]$ ensures the consistency between the reconstructed volumetric image $f(x, y, z)$ and the observation $Y^\theta(u, v)$. The first term, $E_1[f]$, known as TV semi-norm, has been shown to be extremely powerful[41] to remove

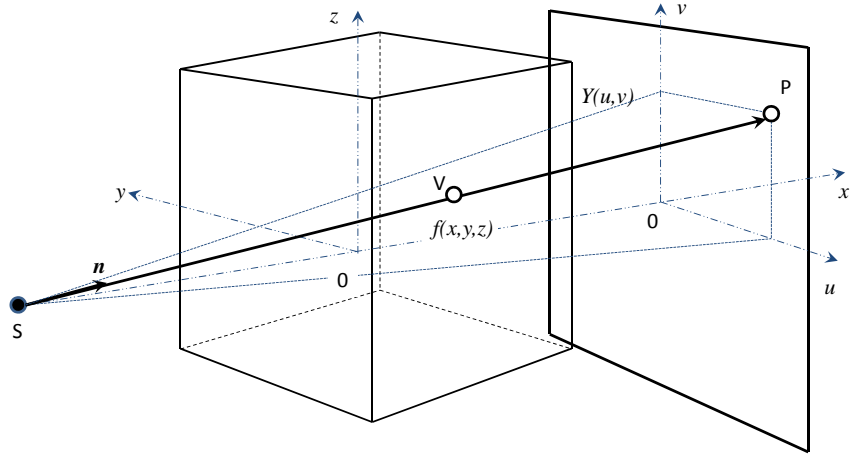

**FIG. 1**. The geometry of x-ray projection. The operator $P^\theta$ maps $f(x,y,z)$ in $\mathbf{R}^3$ onto another function $P^\theta[f(x,y,z)](u,v)$ in $\mathbf{R}^2$, the x-ray imager plane, along a projection angle $\theta$. $L(u,v)$ is the length of SP and $l(x,y,z)$ is that of SV. The source to imager distance is $L_0$.

artifacts and noise from $f$ while preserving its sharp edges to a certain extent. The CBCT image reconstruction from few projections is a well known underdetermined problem in that there are infinitely many functions $f$ such that $P^\theta[f(x,y,z)](u,v) = Y^\theta(u,v)$. By performing the minimization as in Eq. (2) the projection condition will be satisfied to a good approximation. The TV regularization term serves the purpose of picking out the one with desired image properties, namely smooth while with sharp edges, among those infinitely many candidate solutions. The dimensionless constant $\mu > 0$ controls the smoothness of the reconstructed images by adjusting the relative weight between $E_1[f]$ and $E_2[f]$. In the reconstruction results shown in this paper, we choose the value of $\mu$ manually for best image quality.

**B. Minimization Algorithm**

The main obstacle encountered while solving Eq. (2) is due to the projection operator $P^\theta$. In the matrix representation, this operator becomes a sparse matrix. However, it contains approximately $4 \times 10^9$ non-zero elements for a typical clinical case studied in this paper, occupying about 16 GB memory space. This matrix is usually too large to be stored in typical computer memory and therefore it has to be computed repeatedly whenever necessary. This repeated work consumes a large amount of the computational time. For example, if we were to solve Eq. (2) with a simple gradient descent method, $P^\theta$ would have to be evaluated repeatedly while computing the search direction, *i.e.* negative gradient, and while calculating the functional value for step size search as well. This would significantly limit the computation efficiency.

In order to overcome this difficulty, we utilize the forward-backward splitting algorithm[42-44]. This algorithm splits the minimization of the TV term and the data fidelity term into two alternating steps, while the computation of x-ray projection $P^\theta$ is only in one of them. The computation efficiency can be improved owing to this splitting. As such, let us consider the optimality condition of Eq. (2) by setting the functional variation with respect to function $f(x,y,z)$ to zero:

$$\frac{\delta}{\delta f(x,y,z)} E_1[f] + \mu \frac{\delta}{\delta f(x,y,z)} E_2[f] = 0. \tag{3}$$

155    If we split the above equation into the following form

$$\begin{aligned} \mu \frac{\delta}{\delta f(x,y,z)} E_2[f] &= \frac{\beta}{V}(f-g), \\ \frac{\delta}{\delta f(x,y,z)} E_1[f] &= -\frac{\beta}{V}(f-g), \end{aligned} \tag{4}$$

where $\beta > 0$ is a parameter and $g(x,y,z)$ is an auxiliary function used in this splitting algorithm, the minimization problem can be accordingly split, leading to the following iterative algorithm to solve Eq. (2):

---

**Algorithm A1:**

Do the following steps until convergence

1. Update: $g = f - \frac{\mu V}{\beta} \frac{\delta}{\delta f(x,y,z)} E_2[f]$;

2. Minimize: $f = \arg\min E_1[f] + \frac{\beta}{2} E_3[f]$;

3. Correct: $f(x,y,z) = 0$, if $f(x,y,z) < 0$.

---

160

Here a new energy functional is defined as $E_3[f] = \frac{1}{V}\|f-g\|_2^2$. The Step 1 in Algorithm A1 is a gradient descent update with respect to the minimization of energy functional $E_2[f]$. It is also interesting to observe that the Step 2 here is just an Rudin-Osher-Fatemi (ROF) model, which has been shown to be extremely efficient and capable of removing
165    noise and artifacts from a degraded image $g(x,y,z)$ while still preserving the sharp edges and main features[41]. In addition, since $f$ represents the x-ray attenuation coefficients, it is necessary to ensure its non-negativeness by a simple truncation as in Step 3. These three steps are iterated until a solution to the problem in Eq. (2) is achieved.

The choice of $\beta$ is crucial in this algorithm. On one hand, a small value of $\beta$ will lead
170    to a large step size of the gradient descent update in Step 1, causing instability of the forward-backward splitting algorithm. On the other, a large $\beta$ will tend to make the TV semi-norm $E_1[f]$ unimportant in Step 2, reducing the its efficacy in removing artifacts. In practice, an empirical value $\beta \sim 10\mu$ is found to be appropriate.

For the sub-problem in Step 2, optimizing an energy functional $E_{ROF}[f] = E_1[f] +$
175    $\frac{\beta}{2} E_3[f]$, we solve it with a simple gradient descent method. At each iteration step of the gradient descent method, the functional value $E_0 = E_{ROF}[f]$ is first evaluated, as well as the functional gradient $d(x,y,z) = \frac{\delta}{\delta f(x,y,z)} E_{ROF}[f]$. An inexact line search is then performed along the negative gradient direction with an initial step size $\lambda = \lambda_0$. The trial functional value $E_{new} = E_{ROF}[f - \lambda d]$ is evaluated. If Amijo's rule is satisfied[45],
180    namely

$$E[f - \lambda d] \leq E_0 - c\lambda \frac{1}{V} \int \mathrm{d}x \mathrm{d}y \mathrm{d}z \, d(x,y,z)^2 \,, \tag{5}$$

where $c$ is a constant, the step size $\lambda$ is accepted and an update of the image $f \leftarrow f - \lambda d$ is applied; otherwise, the search step size is reduced by a factor $\alpha$ with $\alpha \in (0,1)$. This iteration is repeated until the relative energy decrease in a step $|E_{new} - E_0|/E_0$ is less than a given threshold $\varepsilon$. The parameters relevant in this sub-problem are chosen as empirical values of $c = 0.01$, $\alpha = 0.6$ and $\varepsilon = 0.1\%$. The computation results are found to be insensitive to these choices.

Boundary condition is another important issue during the implementation. For simplicity, zero boundary condition is imposed in our computation along the anterior-posterior direction and the lateral direction, while reflective boundary condition is used along the superior-inferior direction.

### C. Further improvements

#### *1. GPU implementation*

Computer graphic cards, such as the NVIDIA GeForce series and the GTX series, are conventionally used for display purpose on desktop computers. Recently, NVIDIA introduced special GPUs dedicated for scientific computing, for example the Tesla C1060 card that is used in this paper. Such a GPU card has a total number of 240 processor cores (grouped into 30 multiprocessors with 8 cores each), each with a clock speed of 1.3 GHz. The card is also equipped with 4 GB DDR3 memory, shared by all processor cores. Utilizing such a GPU card with tremendous parallel computing ability can considerably elevate the computation efficiency of our algorithm. In fact, a number of components can be easily accomplished in a parallel fashion. For instance, the evaluating of functional value $E_{ROF}[f]$ in Step 2 of Algorithm A1 can be performed by first evaluating its value at each $(x,y,z)$ coordinate and then sum over all coordinates. The functional gradient $d(x,y,z)$ can be also computed with each GPU thread responsible for one coordinate.

#### *2. Closed-form gradient*

A straightforward way of implementing Algorithm A1 is to interpret $P^\theta[f]$ as a matrix multiplication and then $E_2[f]$ as a vector norm $\sum_\theta \|P^\theta f - Y^\theta\|_2^2$. This leads to a simple form $\sum_\theta P^{\theta^T}(P^\theta f - Y^\theta)$ for the functional variation $\delta E_2[f]/\delta f(x,y,z)$ in Step 1, apart from some constants, where $\cdot^T$ denotes matrix transpose. In practice $P^\theta$ is too large to be stored in computer memory. Note that $P^\theta f$ is simply a forward projection calculation, it can be easily computed by a ray-tracing algorithm, such as Siddon's algorithm[46-48]. Due to the massive parallelization ability of GPU, multiple threads can compute projections of a large number rays simultaneously and high efficiency can be achieved. On the other hand, we lack an efficient algorithm to perform the operation of $P^{\theta^T}$. In fact, $P^{\theta^T}$ is a backward operation in that we tend to update voxel values along a ray line. If we were to perform this backward operation by Siddon's algorithm in the GPU implementation with each thread responsible for updating voxels along a ray line, a memory conflict problem would take place due to the possibility of updating a same voxel value by different GPU threads. As a consequence, one thread will have to wait until another thread finishes updating. It is this fact that severely limits the exploitation of GPU's massive parallel computing power.

To overcome this difficulty, we analytically compute the functional variation $\frac{\delta}{\delta f(x,y,z)} E_2[f]$ in Step 1 of Algorithm A1 and a closed-form equation is obtained

$$\frac{\delta}{\delta f(x,y,z)} E_2[f] = \frac{1}{N_\theta A} \sum_\theta \frac{2L^3(u^*,v^*)}{L_0 l^2(x,y,z)} [P^\theta[f(x,y,z)](u^*,v^*) - Y^\theta(u^*,v^*)] . \qquad (6)$$

Here $(u^*, v^*)$ is the coordinate for a point P on imager, where a ray line connecting the source $S = (x_s, y_s, z_s)$ and the point $V = (x, y, z)$ intersects with the imager. $L_0$ is the distance from the x-ray source S to the imager. $l(x,y,z)$ and $L(u^*, v^*)$ are the distance from S to the point V and from S to the point P on imager, respectively. See Fig. 1 for the geometry. The derivation of Eq. (6) is briefly shown in the Appendix.

The form of Eq. (6) suggests a very efficient way of evaluating this functional variation and hence accomplishing Step 1 in Algorithm A1 in a parallel fashion. For this purpose, we can first perform the forward projection operation and compute an auxiliary function defined as $T^\theta(u,v) \equiv [P^\theta[f(x,y,z)](u,v) - Y^\theta(u,v)]$ for all $(u,v)$ and $\theta$. For a point $(x, y, z)$ where we try to evaluate the functional variation, it suffices to take the function values of $T^\theta(u^*, v^*)$ at a $(u^*, v^*)$ coordinate corresponding to the $(x, y, z)$, multiply by proper prefactors, and finally sum over $\theta$. In numerical computation, since we can only evaluate $T^\theta(u,v)$ at a set of discrete $(u,v)$ coordinates and $(u^*, v^*)$ does not necessarily coincide with these discrete coordinates, a bilinear interpolation is performed to get $T^\theta(u^*, v^*)$. Because the computation of $T^\theta(u,v)$ can be achieved in a parallel manner with each GPU thread responsible for a ray line and the evaluation of $\frac{\delta}{\delta f(x,y,z)} E_2[f]$ can be performed with each thread for a voxel $(x, y, z)$, extremely high efficiency in Step 1 is expected given the vast parallelization ability of GPU.

### *3. Multi-grid method*

Another technique we employed to increase computation efficiency is multi-grid method[49]. Because of the convexity of the energy functional in Eq. (2), the minimization problem is equivalent to solving a set of nonlinear differential equations posed by the optimality condition as in Eq. (3). It has long been known that, when solving a differential equation with a certain kind of finite difference scheme, the convergence rate of an iterative approach largely depends on the mesh grid size. In particular, the convergence rate is usually worsened when a very fine grid size is used. Moreover, finer grid also implies more number of unknown variables, significantly increasing the size of the computation task.

A well known multi-grid approach can be utilized to resolve these problems. Suppose it is our ultimate purpose to reconstruct a volumetric CBCT image $f(x, y, z)$ on a fine grid $\Omega_h$ of size $h$, we could start with solving the problem on a coarser grid $\Omega_{2h}$ of size $2h$ with the same iterative approach as in Algorithm A1. Upon convergence, we could smoothly extend the solution $f_{2h}$ on $\Omega_{2h}$ to the fine grid $\Omega_h$ and use it as the initial guess of the solution. Because of the decent quality of this initial guess, only few iteration steps of Algorithm A1 are required to achieve the final solution on $\Omega_h$. Moreover, this two-level idea can be used recursively. In practice, we performed a 4-level multi-grid scheme, *i.e.* the reconstruction is sequentially achieved on grids $\Omega_{8h} \to \Omega_{4h} \to \Omega_{2h} \to \Omega_h$. A considerably efficiency gain is observed in our implementation.

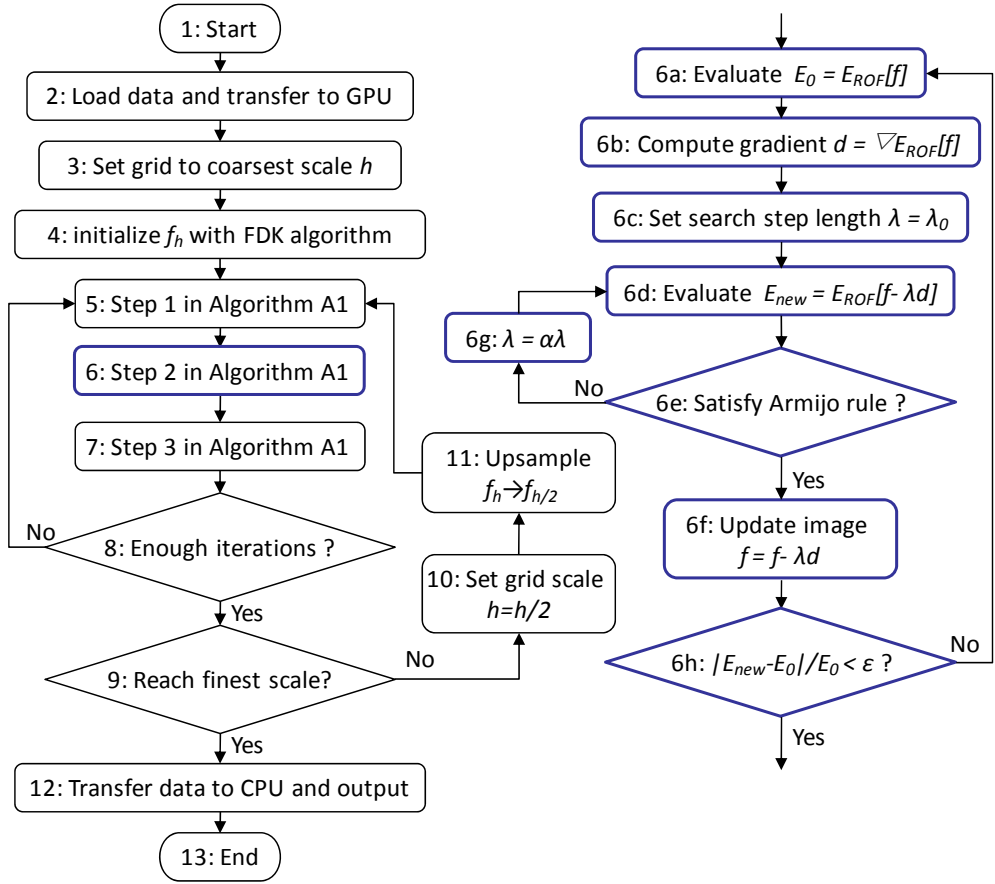

**FIG. 2**. The flow chart of our GPU-based CBCT reconstruction. Step 6 in left panel is enlarged in detail in right panel.

Finally, we summarize our algorithm of CBCT reconstruction, including all techniques mentioned previously, in Fig. 2.

## III. Results

### A. Digital phantom

We test our reconstruction algorithm with a digital NURBS-based cardiac-torso (NCAT) phantom[50], which maintains a high level of anatomical realism (*e.g.*, detailed bronchial trees). The phantom is generated at thorax region with a size of $512 \times 512 \times 70$ voxels and the x-ray imager is modeled to be an array of $512 \times 384$. The source to axes distance is 100 cm and the source to detector distance is 150 cm. X-ray projections are computed along various directions with Siddon's ray tracing algorithm[46-48]. Three slice cuts of the phantom are displayed in Fig. 3, as well as a typical x-ray projection along the AP direction.

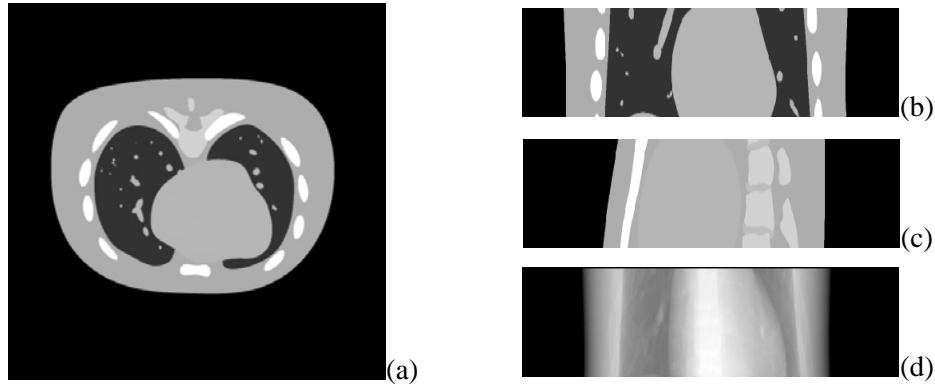

**FIG. 3**. The NCAT phantom used in our CBCT reconstruction is shown in (a) axial view, (b) coronal view, and (c) sagittal view. The x-ray projection along the AP direction is also shown in (d).

We first reconstruct CBCT images based on different number of x-ray projections $N_\theta$. In all cases, the projections are taken along equally spaced angles covering an entire 360 degree rotation. The reconstruction results under $N_\theta = 5$, 10, 20, and 40 projections are listed in Fig. 4. For the purpose of comparison, we also show here the images reconstructed from conventional FDK algorithm[1] with same experimental setting. Clearly, the reconstructed CBCT image from our method based on 40 projections is almost visually indistinguishable from the ground-truth. On the other hand, the one produced by the conventional FDK algorithm is full of streak artifacts due to the insufficient number of projections. Moreover, the required number of projections can be further lowered for some clinical applications. For example, 20 projections may suffice for patient setup purposes in radiotherapy, where major anatomical features have already been retrieved. As far as radiation dose is concerned, the results shown in Fig. 4 implies a 9 times or more dose reduction compared with the currently widely used FDK algorithm, where about 360 projections are usually taken in a full-fan head-and-neck scanning protocol.

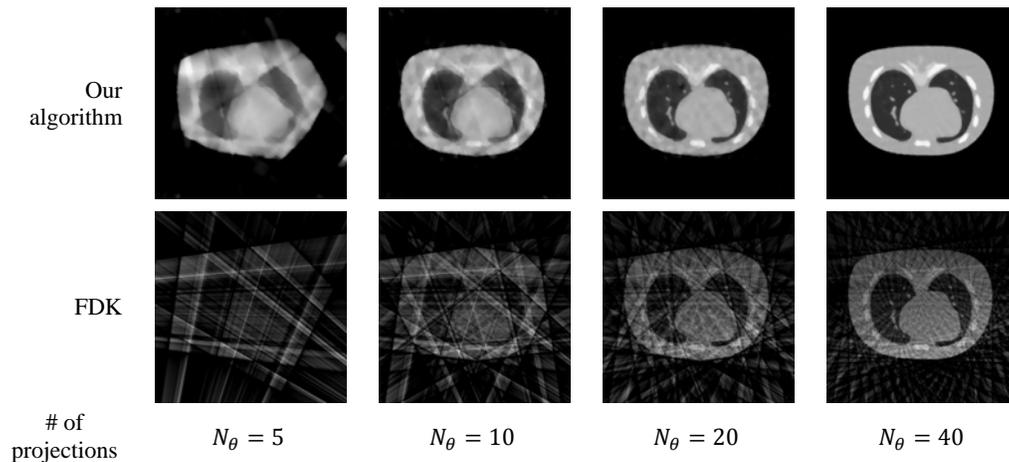

**FIG. 4.** Axial view of the reconstructed CBCT image under $N_\theta = 5$, 10, 20, and 40 x-ray projections are shown. The top row is obtained from our new algorithm, while the bottom one is from the FDK algorithm.

| $N_\theta$ | Error $e$ (%) | Correlation $c$ | Computation Time $t$ (*sec*) |
|---|---|---|---|
| 5  | 28.63 | 0.9386 | **38.7** |
| 10 | 15.96 | 0.9813 | **51.2** |
| 20 | 11.38 | 0.9906 | **77.8** |
| 40 | 7.10  | 0.9963 | **130.3** |

**TABLE. 1:** The relative error $e$, the correlation coefficient $c$, and the computation time $t$ as functions of the number of x-ray projections $N_\theta$ used in our CBCT reconstruction.

In order to quantify our reconstruction accuracy, we use the correlation coefficient $c \equiv \mathrm{corr}(f, f_0)$ and the relative error $e \equiv \|f - f_0\|_2 / \|f_0\|_2$ as metrics to measure the closeness between the ground truth image $f_0(x, y, z)$ and the reconstruction results $f(x, y, z)$. The relative error $e$ and the correlation coefficient $c$ corresponding to the results shown in Fig. 4 are summarized in Table 1. As expected, the more projections used, the better reconstruction quality will be obtained, leading to smaller relative error $e$ and higher correlation coefficient $c$. More importantly, we emphasize here that the total reconstruction time is short enough for real clinical applications. As we can see from Table 1, the reconstructions can be accomplished within 77~130 seconds on a NVIDIA Tesla C1060 GPU card, when 20~40 projections are used. Comparing with the computational time of several hours in currently similar reconstruction approaches, our algorithm has achieved a tremendous efficiency enhancement (~100 times speed up).

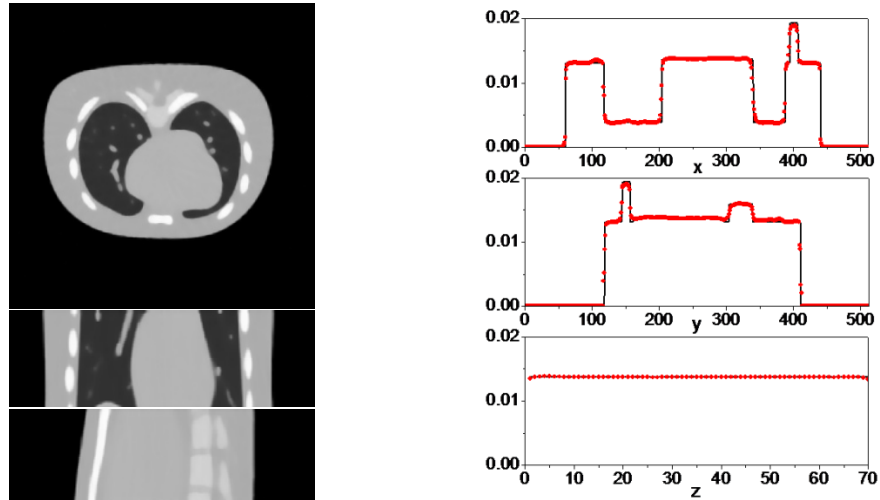

**FIG. 5.** The reconstructed images from 40 projections are shown in left column. From top to bottom are axial, coronal, and sagittal view. In the right column, the image intensity profiles through the centers of the images in three directions are shown in red circles. The corresponding ground-truth profiles are indicated by solid lines.

Finally, we provide a complete visualization of the reconstructed CBCT image in Fig. 5, where three orthogonal planes of the final reconstructed image with $N_\theta = 40$ projections are drawn. To have a clear visual comparison between the reconstructed images and the ground truth, we also plot the image profiles along the central lines in *x*, *y*, and *z* directions. The corresponding profiles in the ground-truth image are indicated by solid lines. These plots clearly show that the reconstructed image, though containing small fluctuations, is close to the ground-truth data to a satisfactory extent.

### B. Catphan phantom

To demonstrate our algorithm's performance in a real physical phantom, we performed CBCT reconstruction on a CatPhan 600 phantom (The Phantom Laboratory, Inc., Salem, NY). 379 projections within 200 degrees are acquired using a Varian On-Board Imager system (Varian Medical Systems, Palo Alto, CA) at 2.56 mAs/projection under a full-fan mode. A subset of equally spaced 40 projections is used to perform the reconstruction. In Fig. 6, we show the reconstruction results of the CatPhan phantom at four axial slices obtained from our algorithm and from the conventional FDK algorithm. Clearly, the images obtained from our method are smooth and major features of the phantom are captured. On the other hand, the FDK algorithm leads to images contaminated by serious streaking artifacts.

To further test the capability of handling noisy data, we have also performed CBCT scans under different mAs levels and reconstructions are then conducted using our TV-based algorithm and the FDK algorithm, as in Fig. 7. Again, the images produced by our method are smooth and free of streaking artifacts, undoubtedly outperforming those from the FDK algorithm. In particular, under an extremely low mAs level of 0.1 mAs/projection, our method is still able to capture major features of the phantom. Comparing with the currently widely used full-fan head-and-neck scan protocol of 0.4 mAs/projection, this performance implies a dose reduction by a factor of ~4 due to lowering the mAs level. Taking into account the dose reduction by reducing the number of x-ray projections, an overall 36 times dose reduction has been achieved.

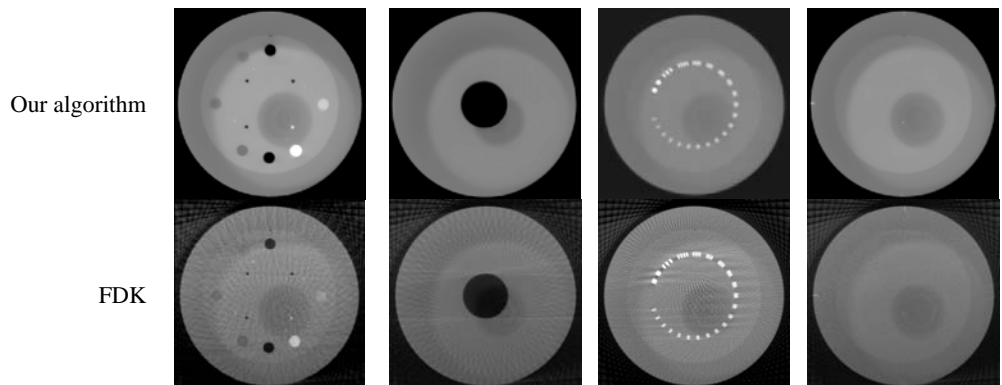

**FIG. 6.** The axial views of the reconstructed images of a CatPhan phantom from $N_\theta = 40$ projections based on our method (top row) and the FDK method (bottom row) at 2.56 mAs/projection. Different columns correspond to different layers at the phantom.

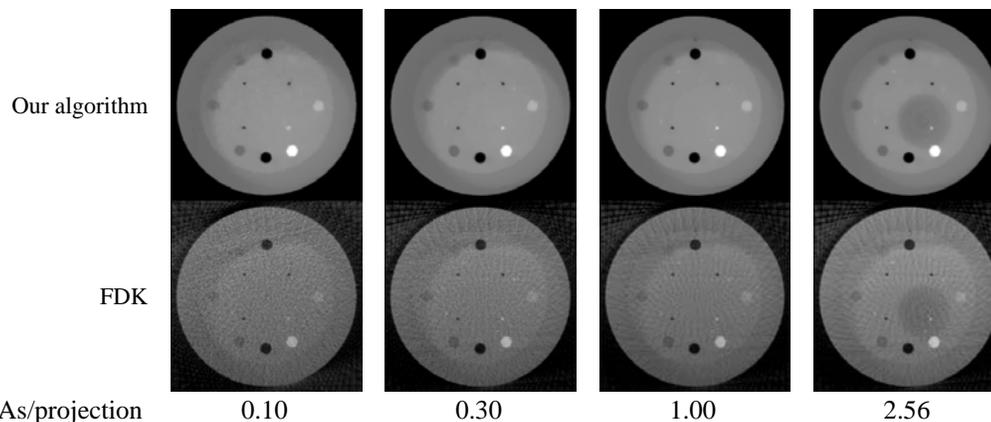

**FIG. 7.** Axial view of the reconstructed CBCT images of a CatPhan phantom at various mAs levels for $N_\theta = 40$ projections.

**C. Head-and-neck Case**

We have also validated our CBCT reconstruction algorithm on realistic head-and-neck anatomical geometry. We take a patient head-and-neck CBCT scan using a Varian On-Board Imager system with 363 projections in 200 degrees and 0.4 mAs/projection. A subset of only 40 projections is selected for the reconstruction. Fig. 8 shows the reconstruction results in this case from our algorithm as well as from the convential FDK algorithm. Due to the complicated geometry and high contrast between bony structures and soft tissues in this head-and-neck region, streak artifacts are extremely severe in the images obtained from FDK algorithm under the undersampling case. In contrast, our algorithm successfully leads to decent CBCT image quality, where artifacts are hardly observed and high image contrast is maintained. In particular, when a metal dental filling exists in the patient, our algorithm can still capture it with high contrast, whereas FDK algorithm produces a number of streaks in the CBCT image.

We have also performed CBCT scans on an anthropomorphic skeleton Rando head phantom (The Phantom Laboratory, Inc., Salem, NY) to validate our algorithm under low mAs levels in such a complicated anatomy. 363 projections within 200 degrees are acquired using a Varian On-Board Imager system at various mAs/projection levels. The CBCT reconstruction uses only a subset of equally spaced 40 projections. In Fig. 9, we show the reconstruction results of this phantom under 0.4 mAs/projection, the current standard scanning protocol for head-and-neck patient. These results are presented in axial view at three different slices as well as in segittal view. In addition, we present the CBCT reconstruction results under different mAs levels ranging from 0.1 mAs/projection to 2.0 mAs/projection in Fig. 10. In all of these testing cases, our method can reconstruct high quality CBCT images, even under low mAs levels at low number of projections. This demonstrates the advantages of our algorithm over the conventional FDK algorithm. As far as the dose reduction, a factor of 36 has been achieved in this head-and-neck case.

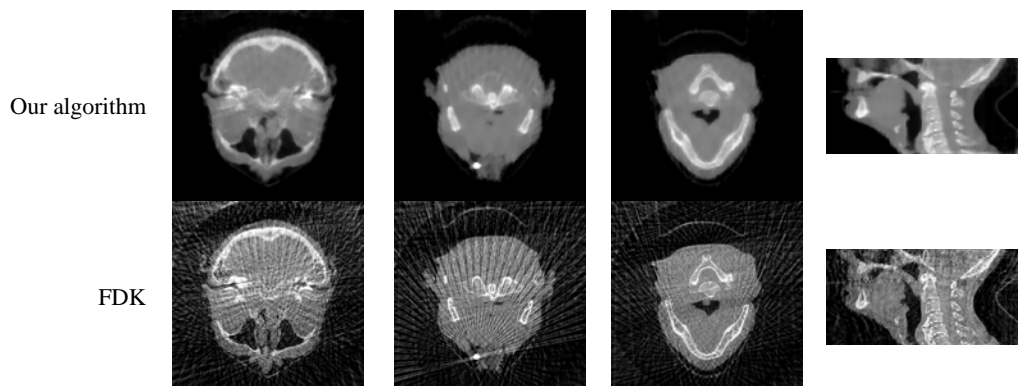

**FIG. 8.** The reconstructed CBCT images of a patient from $N_\theta = 40$ x-ray projections based on our algorithm (top row) and the FDK algorithm (bottom row). The first three columns correspond to axial views at different layers and the last column is sagittal view.

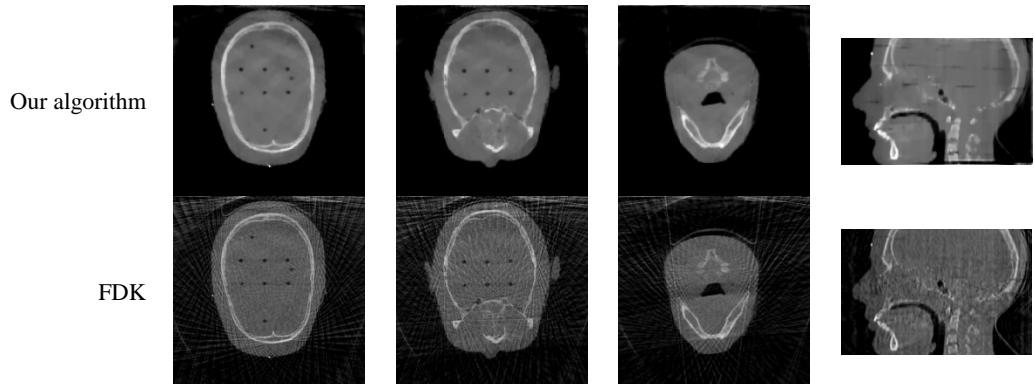

**FIG. 9.** The reconstructed images of a Rando head phantom from $N_\theta = 40$ x-ray projections based on our method (top row) and the FDK method (bottom row). The first three columns correspond to axial views at different layers and the last column is sagittal view. The horizontal dark lines in the segittal view are separations between neighboring slice sections of the phantom.

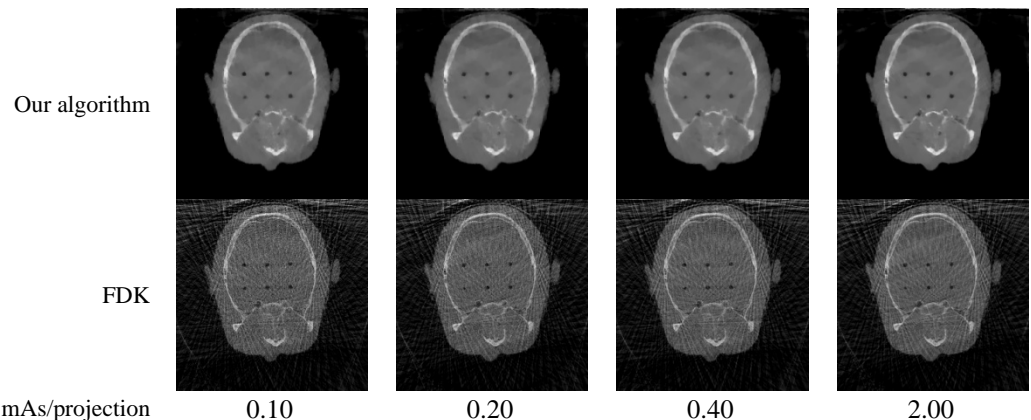

mAs/projection           0.10                0.20               0.40               2.00

**FIG. 10.** Axial view of the reconstructed CBCT images of a head phantom at various mAs levels for $N_\theta = 40$ projections.

## IV. Conclusion

In this paper, we have developed a fast iterative algorithm for CBCT reconstruction. We consider an energy functional consisting of a data fidelity term and a regularization term of TV semi-norm. The minimization problem is solved with a GPU-friendly forward-backward splitting method together with a multi-grid approach on a GPU platform, leading to both satisfactory accuracy and efficiency.

    Reconstructions performed on a digital NCAT phantom and a real patient at the head-and-neck region indicate that images with decent quality can be reconstructed from 40 x-ray projections in about 130 seconds. We have also tested our algorithm on a CatPhan 600 phantom and Rando head phantom under different mAs levels and found that CBCT images can be successfully reconstructed from scans with as low as 0.1

mAs/projection. All of these results indicate that our new algorithm has improved the efficiency by a factor of 100 over existing similar iterative algorithms and reduced imaging dose by a factor of at least 36 compared to the current clinical standard full-fan head and neck scanning protocol. The high computation efficiency achieved in our algorithm makes the iterative CBCT reconstruction approach applicable in real clinical environments.

## Acknowledgement

This work is supported in part by the University of California Lab Fees Research Program. We would like to thank NVIDIA for providing GPU cards for this project.

## Appendix
### Derivation of Eq. (6)

Without losing of generality, assume the x-ray projection is taken along positive x-direction. With the help of delta functions, we could rewrite Eq. (1) as

$$P^\theta[f(x,y,z)](u,v) = \int dl dx dy dz\, f(x,y,z) \cdot \delta(x - x_S - n_1 l)\delta(y - y_S - n_2 l)\delta(z - z_S - n_3 l). \quad (A1)$$

From Fig. 1, the unit vector $\boldsymbol{n} = (n_1, n_2, n_3)$ can be expressed:

$$n_1 = \frac{L_0}{L(u,v)}, \quad n_2 = -\frac{u}{L(u,v)}, \quad n_3 = \frac{v}{L(u,v)}, \quad (A2)$$

where $L(u,v) = [L_0^2 + u^2 + v^2]^{1/2}$. Now for the functional $E_2[f]$ in Eq. (2), we are ready to take variation with respect to $f(x,y,z)$, yielding

$$\frac{\delta}{\delta f(x,y,z)} E_2[f] = \frac{1}{N_\theta A} \sum_\theta 2 \int du dv dl\, [P^\theta[f(x,y,z)](u,v) - Y^\theta(u,v)] \cdot \delta(x - x_S - n_1 l)\delta(y - y_S - n_2 l)\delta(z - z_S - n_3 l). \quad (A3)$$

Let us define functions

$$h_1(u,v,l) = x - x_S - n_1 l = x - x_S - \frac{L_0}{L(u,v)} l,$$
$$h_2(u,v,l) = y - y_S - n_2 l = y - y_S + \frac{u}{L(u,v)} l, \quad (A4)$$
$$h_3(u,v,l) = z - z_S - n_3 l = z - z_S - \frac{v}{L(u,v)} l.$$

From the properties of delta function, it follows that

$$\frac{\delta}{\delta f(x,y,z)} E_2[f] = \frac{1}{N_\theta A} \sum_\theta 2\, [P^\theta[f(x,y,z)](u^*, v^*) - Y^\theta(u^*, v^*)] \frac{1}{\left|\frac{\partial(h_1, h_2, h_3)}{\partial(u,v,l)}\right|_{(u^*,v^*,l^*)}}, \quad (A5)$$

where $(u^*, v^*, l^*)$ is the solution to Eq. (A4). Explicitly, we have

$$u^* = -(y - y_s)L_0/(x - x_s),$$
$$v^* = (z - z_s)L_0/(x - x_s), \quad (A6)$$
$$l^* = L(u^*, v^*)(x - x_s)/L_0.$$

The geometric meaning of this solution is clear. $l^*$ is the distance from x-ray source to the point $V = (x, y, z)$. $(u^*, v^*)$ is the coordinate for a point P on imager, where a ray line connecting the source $S = (x_s, y_s, z_s)$ and the point $V = (x, y, z)$ intersects with the imager. Finally, we need evaluate the Jacobian $\left|\frac{\partial(h_1,h_2,h_3)}{\partial(u,v,l)}\right|$ in Eq. (A5). This somewhat tedious work yields a simple results:

$$\left|\frac{\partial(h_1,h_2,h_3)}{\partial(u,v,l)}\right|_{(u^*,v^*,l^*)} = \frac{L_0 l^2(x,y,z)}{L^3(u^*,v^*)}. \tag{A7}$$

410   Substituting (A7) into (A5) leads to Eq. (6).